\documentclass[superscriptaddress,
showpacs,aps,prd]{revtex4}
\usepackage{amssymb}
\usepackage{epsfig}
  
\begin{document}

\title{\large \bf \boldmath Study of the process $e^+e^-\to\eta\gamma$
in the center-of-mass energy range 1.07--2.00 GeV}
\author{M.~N.~Achasov}
\affiliation{Budker Institute of Nuclear Physics, SB RAS, Novosibirsk, 630090, 
Russia}
\affiliation{Novosibirsk State University, Novosibirsk, 630090, Russia}
\author{V.~M.~Aulchenko}
\affiliation{Budker Institute of Nuclear Physics, SB RAS, Novosibirsk, 630090, 
Russia}
\affiliation{Novosibirsk State University, Novosibirsk, 630090, Russia}
\author{A.~Yu.~Barnyakov}
\affiliation{Budker Institute of Nuclear Physics, SB RAS, Novosibirsk, 630090, 
Russia}
\author{K.~I.~Beloborodov}
\author{A.~V.~Berdyugin}
\email[e-mail:]{berdugin@inp.nsk.su}
\affiliation{Budker Institute of Nuclear Physics, SB RAS, Novosibirsk, 630090, 
Russia}
\affiliation{Novosibirsk State University, Novosibirsk, 630090, Russia}
\author{A.~G.~Bogdanchikov}
\author{A.~A.~Botov}
\affiliation{Budker Institute of Nuclear Physics, SB RAS, Novosibirsk, 630090, 
Russia}
\author{T.~V.~Dimova}
\author{V.~P.~Druzhinin}
\author{V.~B.~Golubev}
\author{K.~A.~Grevtsov}
\author{L.~V.~Kardapoltsev}
\author{A.~G.~Kharlamov}
\affiliation{Budker Institute of Nuclear Physics, SB RAS, Novosibirsk, 630090, 
Russia}
\affiliation{Novosibirsk State University, Novosibirsk, 630090, Russia}
\author{D.~P.~Kovrizhin}
\affiliation{Budker Institute of Nuclear Physics, SB RAS, Novosibirsk, 630090, 
Russia}
\author{I.~A.~Koop}
\author{A.~A.~Korol}
\affiliation{Budker Institute of Nuclear Physics, SB RAS, Novosibirsk, 630090, 
Russia}
\affiliation{Novosibirsk State University, Novosibirsk, 630090, Russia}
\author{S.~V.~Koshuba}
\author{A.~P.~Lysenko}
\author{K.~A.~Martin}
\author{A.~E.~Obrazovsky}
\author{E.~V.~Pakhtusova}
\affiliation{Budker Institute of Nuclear Physics, SB RAS, Novosibirsk, 630090, 
Russia}
\author{E.~A.~Perevedentsev}
\affiliation{Budker Institute of Nuclear Physics, SB RAS, Novosibirsk, 630090, 
Russia}
\affiliation{Novosibirsk State University, Novosibirsk, 630090, Russia}
\author{A.~L.~Romanov}
\affiliation{Budker Institute of Nuclear Physics, SB RAS, Novosibirsk, 630090, 
Russia}
\author{S.~I.~Serednyakov}
\author{Z.~K.~Silagadze}
\affiliation{Budker Institute of Nuclear Physics, SB RAS, Novosibirsk, 630090, 
Russia}
\affiliation{Novosibirsk State University, Novosibirsk, 630090, Russia}
\author{A.~N.~Skrinsky}
\author{I.~K.~Surin}
\affiliation{Budker Institute of Nuclear Physics, SB RAS, Novosibirsk, 630090, 
Russia}
\author{Yu.~A.~Tikhonov}
\author{A.~V.~Vasiljev}
\affiliation{Budker Institute of Nuclear Physics, SB RAS, Novosibirsk, 630090,
Russia}
\affiliation{Novosibirsk State University, Novosibirsk,
630090, Russia}
\author{P.~Yu.~Shatunov}
\affiliation{Budker Institute of Nuclear Physics, SB RAS, Novosibirsk, 630090,
Russia}
\author{Yu.~M.~Shatunov}
\affiliation{Budker Institute of Nuclear Physics, SB RAS, Novosibirsk, 630090,
Russia}
\affiliation{Novosibirsk State University, Novosibirsk,
630090, Russia}
\author{D.~A.~Shtol}
\affiliation{Budker Institute of Nuclear Physics, SB RAS, Novosibirsk, 630090,
Russia}


\begin{abstract}
The $e^+e^-\to\eta\gamma$ cross section has been measured
in the center-of-mass energy range 1.07--2.00~GeV using
the decay mode $\eta\to 3\pi^0$, $\pi^0\to \gamma\gamma$.
The analysis is based on 36~pb$^{-1}$ of integrated luminosity collected 
with the SND detector at the VEPP-2000 $e^+e^-$ collider.
The measured cross section of about 35 pb at 1.5 GeV is explained
by decays of the $\rho(1450)$ and $\phi(1680)$ resonances.
\end{abstract}

\maketitle
\section{Introduction}
Radiative decays are a powerful tool for studying the internal
structure of hadrons. For light vector mesons, these decays have been 
investigated in several experiments over the past 40 years. 
The probabilities of the $\rho$, 
$\omega$ and $\phi$ decays to $\eta\gamma$ are currently 
measured with accuracies of 7\%, 9\% and 2\%, respectively. For the
$\rho$ and $\omega$ mesons, the errors are still dominated by statistics.
The most accurate measurements of the light vector meson decays to 
$\eta\gamma$ were performed in the SND~\cite{SNDetagam} and 
CMD-2~\cite{CMDetagam} experiments at the VEPP-2M $e^+e^-$ collider. 
These measurements will be continued with more statistics at the 
VEPP-2000 collider~\cite{VEPP2000}.

In $e^+e^-$ experiments a directly measured quantity is the cross section for
$e^+e^-\to\eta\gamma$. The cross section is measured in 
a wide range of the center-of-mass (c.m.) energies, 
for example, from 0.6 to 1.4 GeV
at VEPP-2M~\cite{SNDetagam,CMDetagam}. The decay probabilities are then derived
from the fit to the cross-section data with a sum of vector-resonance 
contributions. When analyzing the VEPP-2M data,
it was found that the model errors on the probabilities of the 
$\rho$, $\omega$, $\phi\to \eta\gamma$ decays associated with uncertainties 
of contributions of excited vector states, can reach several percents. 
To diminish this uncertainty the measurement of the cross 
section for $e^+e^-\to\eta\gamma$ is required at energies at least up to 2~GeV.

Measurement in the 1--2~GeV energy range is also interesting in itself. From 
the cross section data we can derive the probabilities of radiative decays 
of excited vector mesons, such as the $\rho(1450)$ and $\phi(1680)$. In this 
energy region, besides the normal $q\bar{q}$ vector states, production of 
exotic hybrid (quark-antiquark-gluon ) mesons is expected. Since hybrid states
can be mixed with the conventional quark-antiquark states, their identification
is a difficult experimental problem requiring a detailed analysis of all 
possible decay modes. Radiative decays, the probabilities of which are 
expected to be relatively well predicted in the framework of the quark model,
may play the key role in the identification of the hybrid vector states.

In this article we present a measurement of the $e^+e^-\to\eta
\gamma$ cross section in the energy range 1.07--2.00~GeV in an experiment 
with the SND detector at the VEPP-2000 $e^+e^-$ collider~\cite{VEPP2000}.

\section{Detector and experiment}
We analyze data with an integrated luminosity of about 40~pb$^{-1}$ 
accumulated in 2010-2012. During the experiments, the energy range 
1.05-2.00~GeV was scanned several times with a step of 20-25~MeV. In this 
analysis, because of the low statistics, 
we measure the cross section values averaged over ten energy intervals
listed in Table~\ref{tabl2}.

A detailed description of the SND detector is given in Ref.~\cite{SND}. 
This is a nonmagnetic detector, the main part of which is a three-layer 
spherical electromagnetic calorimeter based on NaI(Tl) crystals. A solid angle 
covered by the calorimeter is 90\% of 4$\pi$. Its energy resolution for 
photons is $\sigma_E/E=4.2\%/\sqrt[4]{E({\rm GeV})}$, and the angular 
resolution about $1.5^\circ$. Directions of charged particles are measured 
in the tracking system consisting of a nine-layer drift chamber and 
a proportional chamber with readout from cathode strips. The 
drift chamber provides solid angle coverage of 94\% of 4$\pi$.

The process $e^+e^-\to\eta\gamma$ is studied in the decay mode $\eta\to 3\pi^0
\to 6\gamma$. Since the final state under study does 
not contain charged particles, for normalization we choose 
the process without charged particles, $e^+e^-\to \gamma\gamma$. 
As a result of such normalization, systematic uncertainties associated with 
the event selection in the hardware first-level trigger are canceled, as well 
as uncertainties arising from superimposing beam-generated spurious tracks
onto the events being studied. The uncertainty on the luminosity measurement 
with the process $e^+e^-\to \gamma\gamma$ is estimated to be 
2.2\%~\cite{SNDomegapi}.

\section{Event Selection}
\begin{figure}
\includegraphics[width=0.9\textwidth]{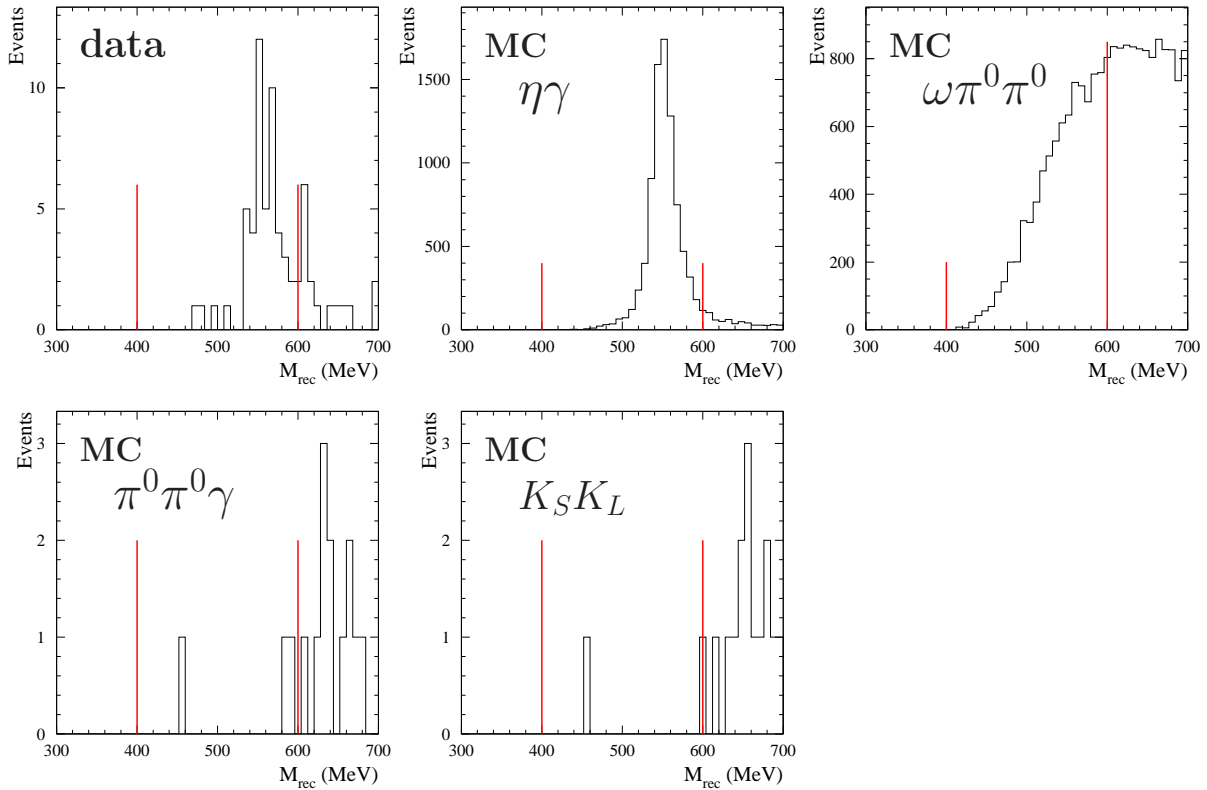}
\caption{ \label{rmg}
The $M_{rec}$ distributions for data  and 
simulation of the signal and background
processes. The vertical lines indicate the selection boundaries.
}
\end{figure}
The main decay modes of the $\eta$ meson are $2\gamma$ (39\%), $3\pi^0$ (33\%) 
and $\pi^+\pi^-\pi^0$ (23\%). Background from the processes $e^+e^-\to 
3\gamma$ and $e^+e^-\to \pi^+\pi^-2\pi^0$ significantly exceeds the
signal in the energy range 1.07--2.00~GeV and does not allow to use the
decay modes $\eta\to 2\gamma$ and $\eta\to \pi^+\pi^-\pi^0$. Thus,
in this paper, the process $e^+e^-\to\eta\gamma$ is studied in the decay 
channel $\eta\to 3\pi^0~,~\pi^0\to 2\gamma$ having seven photons in the final 
state.

The main sources of background in this analysis are the processes 
$e^+e^-\to K_SK_L(\gamma)$ with $K_S\to 3\pi^0$,
$e^+e^-\to\pi^0\pi^0\gamma$ and
$e^+e^-\to\omega\pi^0\pi^0~,~\omega\to\pi^0\gamma$,
of which only the latter has seven photons in the final 
state. In the process  $e^+e^-\to K_SK_L$, additional spurious 
photons originate from $K_L$ nuclear interactions in the calorimeter. 
In the process $e^+e^-\to \omega\pi^0\to\pi^0\pi^0\gamma$, extra photons
can be reconstructed due to splitting of the electromagnetic showers,
photon emission by the initial particles at a large angle, and  
superimposing beam-generated background.

Event selection is carried out in two stages. 
At the first stage we select events containing at least
seven photons and no charged particles. The events must satisfy the 
following conditions on the total energy deposition in the calorimeter 
($E_{tot}$) and the total momentum of photons ($P_{tot}$):
\begin{equation}
0.7 < E_{tot}/2E_{beam} < 1.2,~cP_{tot}/2E_{beam} < 0.3,~
E_{tot}/2E_{beam} - cP_{tot}/2E_{beam} > 0.7.
\end{equation}
At the second stage, kinematic fits are performed for selected events with 
requirements of energy-momentum conservation and $\pi^0$ mass constraints. 
As a result of the kinematic fit, we obtain corrected photon energies and 
$\chi^2$ for the kinematic hypothesis used. Two hypotheses are tested:
\begin{itemize}
\item $e^+e^-\to 3\pi^0\gamma$ ($\chi^2_{3\pi^0\gamma}$),
\item $e^+e^-\to\pi^0\pi^0\gamma$ ($\chi^2_{\pi^0\pi^0\gamma}$).
\end{itemize}
Under the $e^+e^-\to 3\pi^0\gamma$ hypothesis, it is assumed that the recoil 
photon is the most energetic in an event, and $\pi^0$'s are constructed from
the remaining six photons. 
When photons additional  within the tested hypothesis are present 
in the event, we check all possible five(seven)-photon combinations and use
the one with the minimal value of $\chi^2_{\pi^0\pi^0\gamma}$ 
($\chi^2_{3\pi^0\gamma}$). Further selection uses the following conditions:
\begin{equation}
\chi^2_{3\pi^0\gamma}<50,~\chi^2_{\pi^0\pi^0\gamma}>20.
\end{equation}
Under the $e^+e^-\to 3\pi^0\gamma$ kinematic hypothesis, the mass
recoiling against the photon $M_{rec}$ is calculated. The $M_{rec}$ 
distributions for data  as well as for simulation of the process
under study and the background processes are shown in Fig.~\ref{rmg}.
It is seen that the signal process $e^+e^-\to\eta\gamma$ dominates in the 
data distribution. For the final event selection, 
the condition $400<M_{rec}<600$ MeV/$c^2$ is used.

\begin{figure}
\includegraphics[width=0.5\textwidth]{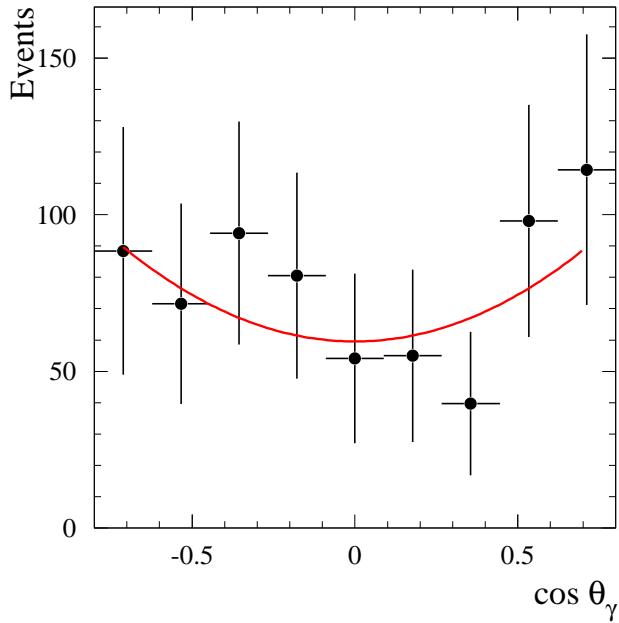}
\caption{ \label{theta}
The efficiency-corrected distribution of the cosine of the recoil-photon 
polar angle for selected data events. The curve 
is a fit to data with $A(1+\cos^2{\theta_{\gamma}})$.
}
\end{figure}
Figure~\ref{theta} shows the $\cos{\theta_\gamma}$ distribution, where 
$\theta_\gamma$ is the recoil-photon polar angle, for selected 
data events with  $\cos{\theta_\gamma}<0.8$. The distribution is 
corrected to take into account the angular dependence of the 
detection efficiency. It is seen that the data are well 
described by the distribution $1+\cos^2{\theta_\gamma}$ expected for
$e^+e^-\to\eta\gamma$.

With the criteria described above, 60 events are selected. Their 
distribution over the energy intervals together with the expected 
background distribution is given in Table~\ref{tabl2}. The background is 
estimated from MC simulation using the measured $\phi\to K_S K_L$ decay
probability~\cite{pdg} and cross sections for the processes 
$e^+e^-\to K_S K_L$~\cite{SNDkskl}, 
$e^+e^-\to \omega\pi^0\to \pi^0\pi^0\gamma$~\cite{SNDomegapi} and 
$e^+e^-\to \omega\pi^+\pi^-$~\cite{BABARomegapipi}. 
For the process $e^+e^-\to \omega\pi^0\pi^0$, we use the isotopic relation
$\sigma(\omega\pi^+\pi^-)=2\sigma(\omega\pi^0\pi^0)$. 
when calculating the background contribution.
Our MC simulation takes into account radiative corrections~\cite{radcor}. 
This is particularly important for the background from the process 
$e^+e^-\to K_S K_L(\gamma)$, which is dominated by
radiative return to the $\phi$ meson through the reaction
$e^+e^-\to \phi\gamma$. The estimated number of background events is equal
to 2.3 (0.4 from $K_S K_L(\gamma)$, 0.5 from  $\pi^0\pi^0\gamma$ and 1.4 from 
$\omega\pi^0\pi^0$). The process $e^+e^-\to K_S K_L(\gamma)$ contributes 
only to the first interval. We conservatively estimate the systematic 
uncertainty in the background calculation to be  
100\% of the background calculated.

\section{Detection efficiency}
\begin{figure}
\includegraphics[width=0.9\textwidth]{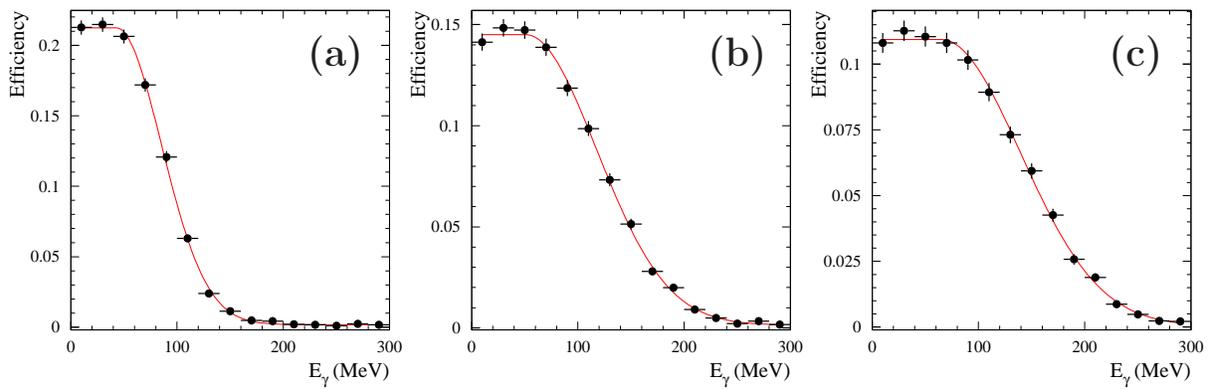}
\caption{ \label{eff}
The detection efficiency for $e^+e^-\to \eta\gamma$ events as a function
of the energy of the additional photon emitted by initial particles for 
(a) $2E_{beam} = 1.1$ GeV, (b) $2E_{beam} = 1.6$ GeV, and (c) 
$2E_{beam} = 1.96$ GeV. The points with 
error bars are obtained using the MC simulation, the curve is the result of 
the $\varepsilon(E_r)$ approximation  by a smooth function.
}
\end{figure}
The detection efficiency for the process under study is determined using MC 
simulation, which takes into account the initial-state radiative 
corrections~\cite{radcor}, in particular, emission of additional photons. 
The angular distribution of these photons is modeled 
according to Ref.~\cite{BM}.

The detection efficiency is determined as a function of two parameters: 
the c.m. energy and the energy of the additional photon $E_r$.
Figure~\ref{eff} shows the dependence of the detection efficiency 
on $E_r$ for three different values of the c.m. energy.
The values of the detection efficiency at $E_r=0$, averaged over the 
corresponding energy intervals, are listed in the Table~\ref{tabl2}.

\section{Fitting the visible cross section and extraction of the Born 
cross section\label{fit}}
The visible cross section for $e^+e^-\to \eta\gamma$ directly 
obtained from the experimental data ($\sigma_{vis} = (N-N_{bkg})/IL$)
is related to the Born cross section ($\sigma(E)$) by the expression:
\begin{equation}
\label{viscrs}
\sigma_{vis}(E) = \int\limits_{0}^{x_{max}} \epsilon_r(E,\frac{xE}{2}) 
F(x,E) \sigma(\sqrt{1-x}E)dx~,
\end{equation}
where $F(x,E)$ is a function~\cite{radcor} describing the distribution of 
the energy fraction, $x=2E_r/E$, carried out by photons emitted from the 
initial state. Equation (\ref{viscrs}) can be rewritten in the traditional 
form:
\begin{equation}
\label{viscrs1}
\sigma_{vis}(E) = \epsilon(E)\,\sigma(E)\,(1+\delta(E)),
\end{equation}
where the detection efficiency $\epsilon(E)$ and the radiative correction 
$\delta(E)$ are defined as follows:
\begin{equation}
\epsilon(E) \equiv \epsilon_r(E,0),
\end{equation}
\begin{equation}
\delta(E) = \frac{\int\limits_{0}^{\frac{2E_{r, max}}{E}}\epsilon_r
(E,\frac{xE}{2}) F(x,E) \sigma(\sqrt{1-x}E)dx }{\epsilon_r(E,0) \cdot 
\sigma(E)} - 1 .
\label{rad}
\end{equation}
The Born cross section is determined as follows. The energy dependence of 
the measured visible cross section is fit with Eq.~(\ref{viscrs}), in which 
the Born cross section is parametrized by a theoretical model that 
describes data reasonably well. The fitted model parameters are used 
to calculate the radiative correction according to Eq.~(\ref{rad}). The 
experimental values of the Born cross section are then obtained
using Eq.~(\ref{viscrs1}).

The energy dependence of the Born cross section for  
$e^+e^-\to \eta\gamma$ is parametrized according to the 
vector meson dominance (VMD) model:
\begin{equation}
\label{parcrs}
\sigma_{\eta\gamma}(E) = \frac{k_\gamma(E)^3}{E^3} \left|
\sum\limits_{V=\rho,~\omega,~\phi, {\ldots} } A_{V}(E)\right|^2,\;\;\;
A_V(E) = \frac{m_V \Gamma_V(m_V) e^{i\varphi_V}}{D_V(E)} \sqrt{ \frac{m^3_V}
{k_\gamma(m_V)^3} \sigma_{V\eta\gamma}},
\end{equation}
\begin{equation}
\label{parcrs1}
D_V(E) = m^2_V -E^2 - i E\Gamma_V(E),\;\;\; k_\gamma(E) = \frac{E}{2} 
\left( 1 - \frac{m^2_{\eta}}{E^2} \right),
\end{equation}
where $E=2E_{beam}$, $m_V$ is the mass of the vector resonance $V$, 
$\Gamma_V(E)$ is its energy-dependent total width, 
$\sigma_{V\eta\gamma} =(12\pi/m_V^2)B(V\to e^+e^-)B(V\to \eta\gamma)$ is 
the cross section for the process $e^+e^-\to V\to\eta\gamma$ at $E=m_V$, 
$B(V\to e^+e^-)$ and $B(V\to \eta\gamma)$ are the probabilities of the 
corresponding decays, $\varphi_V$ is the resonance phase 
($\varphi_{\rho} \equiv 0$).
Besides $\rho$, $\omega$ and $\phi$ resonances, the sum includes all their 
excited states.

In the fit, parameters of the $\rho$, $\omega$ and $\phi$ resonances are fixed
at the nominal values from the PDG tables~\cite{pdg}. The phases of the $\rho$,
$\omega$ and $\phi$ contributions are chosen according to the quark model 
predictions: $\varphi_{\omega}=\varphi_{\rho}$, 
$\varphi_{\phi}=\varphi_{\rho}+180^\circ$.
At energy above 1~GeV the excited vector states $\omega(1420)$, $\rho(1450)$,
$\omega(1650)$, $\phi(1680)$ and $\rho(1700)$ contribute to the 
$e^+e^-\to \eta\gamma$ cross section.
Separation of these resonances in our fit is impossible.
However, we can simplify the problem using the fact that the resonances 
are divided into two groups with similar masses, namely ($\omega(1420)$, 
$\rho(1450)$) and ($\omega(1650)$, $\phi(1680)$, $\rho(1700)$). 
With low  statistics
available we can use a fit model with two effective resonances 
$\rho^{\prime}$ and $\phi^\prime$ with masses and widths equal to the 
PDG values for $\rho(1450)$ and $\phi(1680)$~\cite{pdg}. Such a choice of 
resonances is consistent with predictions of the quark model~\cite{thpred}, 
in which the decay widths of $\rho(1450)\to\eta\gamma$ and $\phi(1680)\to\eta
\gamma$ are at least an order of magnitude larger than the 
corresponding widths for the 
three remaining excited states. The total widths of the $\rho^{\prime}$ and 
$\phi^\prime$ in the formula (\ref{parcrs1}) are assumed to be 
independent of energy.

\begin{figure}
\includegraphics[width=0.5\textwidth]{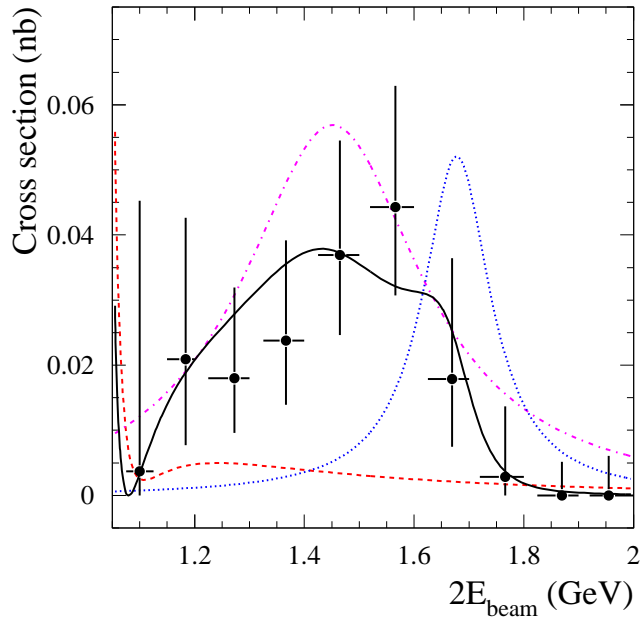}
\caption{ \label{crs1}
The $e^+e^-\to\eta\gamma$ cross section measured in this work. The solid curve
shows the result of the fit with the contributions of the  
$\rho$, $\omega$, $\phi$, $\rho^{\prime}$ and $\phi^{\prime}$ resonances. 
The calculated 
cross section for $e^+e^-\to \rho,\omega,\phi \to \eta\gamma$
is shown separately (dashed curve) as well as the cross sections for 
$e^+e^-\to\rho^\prime\to \eta\gamma$ (dot-dashed curve) and
$e^+e^-\to\phi^\prime\to\eta\gamma$ (dotted curve). Two latter curves are 
calculated using parameters obtained in the fit.
}
\end{figure}

The cross sections $\sigma_{\rho^{\prime}\eta\gamma}$ and $\sigma_{\phi^\prime
\eta\gamma}$ are free fit parameters.
For the phases $\varphi_{\rho^\prime}$ and $\varphi_{\rho^{\prime\prime}}$ 
their canonical values~\cite{clegg} $\varphi_{\rho}+180^\circ$ and 
$\varphi_{\phi}+180^\circ$ are taken.  The fit result is 
shown in Fig.~\ref{crs1} together with the values of the Born cross section
calculated using Eq.~(\ref{viscrs1}). The numerical values of the Born 
cross section and radiative correction are listed in Table~\ref{tabl2}.
It should be noted that almost all data events in the first
energy interval arise from the radiative return process $e^+e^-\to 
\phi\gamma$ and are actually background events.

The fitted values of the cross sections at the resonance peaks are 
following:
\begin{eqnarray}
\sigma_{\rho^{\prime}\to\eta\gamma} = 57 \pm 10 \pm 7\mbox{ pb},\nonumber \\
\sigma_{\phi^{\prime}\to\eta\gamma} = 52 \pm 17 \pm 15\mbox{ pb}. \label{res1}
\end{eqnarray}
The first error is statistical, the second systematic. The systematic
errors were determined by varying the masses and widths of the excited 
resonances within the uncertainties of these parameters for $\rho(1450)$ 
and $\phi(1680)$. Figure~\ref{crs1} shows the cross sections of the processes
$e^+e^-\to \rho^\prime  \to \eta\gamma$ and 
$e^+e^-\to \phi^\prime  \to \eta\gamma$, which correspond to the measured
$\sigma_{\rho^{\prime}\to\eta\gamma}$ and $\sigma_{\phi^{\prime}\to\eta\gamma}$,
and tails from the decays of $\rho$, $\omega$ and $\phi$ mesons, i.e. 
the $e^+e^-\to \rho,\omega,\phi \to \eta\gamma$ cross section.
The fit results make it clear that the 
measured cross section cannot be successfully described without the 
contributions of excited vector mesons.

\section{Systematic uncertainty of the measurement}
The systematic uncertainty on the measured cross section includes 
uncertainties in the detection efficiency determination,
in the luminosity measurement, in the background estimation, and
the model error in the radiative correction calculation.

To estimate the systematic uncertainty on the detection efficiency,
we vary the selection criteria, in particular the condition on 
$\chi^2_{3\pi^0\gamma}$, within wide ranges and 
study the stability of the cross section results. We also perform
the analysis with the requirement of detection of exactly seven photons
in an event. At the existing level of statistical accuracy (60 detected 
events under the standard selection), no change of the cross section results 
is observed. To obtain numerical estimation of the uncertainty on the 
detection efficiency, we use the results of Ref.~\cite{SNDomegapi},
where differences in the detector response between 
data and simulation were studied for the five-photon final state. Using 
much larger statistics, a correction to the detection 
efficiency determined from MC simulation was found to be $(-1.8\pm 1.2)\%$.
For current analysis, a sum of this correction and its error (3\%) is
taken as estimate of the uncertainty on the detection efficiency.

The luminosity is  measured by using events of the two-photon annihilation 
with an accuracy of 2.2\%. The systematic error in the number of selected 
signal events due to background subtraction is estimated to be equal to the 
number of background events.

To estimate the model error in the calculation of the radiative 
correction, we vary within the errors the masses and widths of the 
$\rho^\prime$ and $\phi^\prime$ resonances. The largest effect comes, however, 
from variation of the phase of the $\rho^\prime$ amplitude
which leads to a decrease of the dip in the cross section near 1.07~GeV. The 
change in the cross section due to variation resonance parameters 
and phases reaches 20~pb in the first energy interval and does not exceed 
2~pb in the other. These values are taken as estimates of the model error.
The numerical values of the total systematic uncertainties are listed in 
Table~\ref{tabl2}.  
\begin{table*}
\caption{\label{tabl2}
The energy interval, integrated luminosity ($IL$), number of selected 
events ($N$), estimated number of background events ($N_{bkg}$),  
detection efficiency ($\epsilon_0$), radiative correction ($\delta+1$), 
$e^+e^-\to\eta\gamma$ Born cross section ($\sigma$). The first 
error in the cross section is statistical, the second systematic. For
the last two energy intervals, the upper limits at the 90\% confidence level
are listed for the cross section.}
\begin{ruledtabular}
\begin{tabular}{ccccccc}
$2E_{beam}$ (MeV) & $IL$ (nb$^{-1}$) & $N$ & $N_{bkg}$ & $\epsilon_0$ (\%) 
& $\delta+1$ & $\sigma$ (pb) \\ \hline \\[-2.1ex]
1075--1125  & 1962 & 25  & 0.4  & 7.43  &  45.8 & $ 4^{+41}_{-4}\pm 20$\\
1150--1200  & 2093 &  4  & 0.1  & 6.94  &  1.28 & $21^{+22}_{-13}\pm 2$\\
1225--1300  & 3250 &  4  & 0.2  & 6.99  &  0.93 & $18^{+14}_{- 8}\pm 2$\\
1325--1400  & 3367 &  5  & 0.2  & 6.52  &  0.92 & $24^{+15}_{-10}\pm 2$\\
1425--1500  & 3600 &  8  & 0.3  & 6.23  &  0.93 & $37^{+18}_{-12}\pm 2$\\
1520--1600  & 4351 & 10  & 0.5  & 5.26  &  0.94 & $44^{+19}_{-14}\pm 2$ \\
1625--1700  & 3016 &  3  & 0.3  & 5.21  &  0.96 & $18^{+19}_{-10}\pm 2$ \\
1720--1800  & 4675 &  1  & 0.2  & 4.53  &  1.34 & $ 3^{+11}_{- 3}\pm 1$  \\
1825--1900  & 5134 &  0  & 0.1  & 4.34  &  1.64 & $< 6$           \\
1920--2000  & 4917 &  0  & 0.0  & 3.84  &  1.76 & $< 7$           \\ 
\end{tabular}
\end{ruledtabular}
\end{table*}

\section{\boldmath Comparison with the data on the cross sections of $e^+e^-\to V\eta$,
$V=\rho,\omega,\phi$\label{VMDfit}}
Contributions to the $e^+e^-\to \eta\gamma$ cross section of the isovector 
and isoscalar $n\bar{n}$ ($n=u,d$) and $s\bar{s}$ states can be estimated 
from the cross sections for $e^+e^-\to\rho\eta$, $e^+e^-\to\omega\eta$ and 
$e^+e^-\to\phi\eta$, respectively, within the framework of the VMD model 
as follows:
\begin{equation}
\label{VMD}
\sigma_{\eta\gamma}(E) = \frac{4\pi\alpha}{f_V^2} \sigma_{V\eta} 
\frac{k^3_\gamma}{k^3_V},
\end{equation}
where $f_V$ is the  vector-meson--photon coupling constant, which is 
calculated from the vector meson electronic width: $f_V^2=4\pi m_V\alpha^2/
(3\Gamma(V\to e^+e^-))$. The vector meson momentum $k_V$ for relatively 
narrow $\omega$ and $\phi$ resonances can be calculated as 
\begin{equation}
k_V = \frac{1}{2E}\sqrt{ ((E+m_{\eta})^2 - m^2_V)\cdot 
((E-m_{\eta})^2 - m^2_V)}.
\end{equation}
For the $\rho$-meson one should use the exact formula that takes into account 
the finite width of the resonance~\cite{thetarho}.

Accuracy of Eq.~(\ref{VMD}) can be estimated from experimentally
well-studied similar processes, for example, by comparing the measured decay 
width $\omega\to\pi^0\gamma$  with its VMD estimate from the decay
$\omega\to\rho\pi\to \pi^+\pi^-\pi^0$ (see, for example, \cite{SND2pi0gamma}).
The estimate turns out to be 1.5 times larger than the actual width. One can
therefore conclude that the typical accuracy of 
Eq.~(\ref{VMD}) is about 50\%.
\begin{figure}
\includegraphics[width=0.5\textwidth]{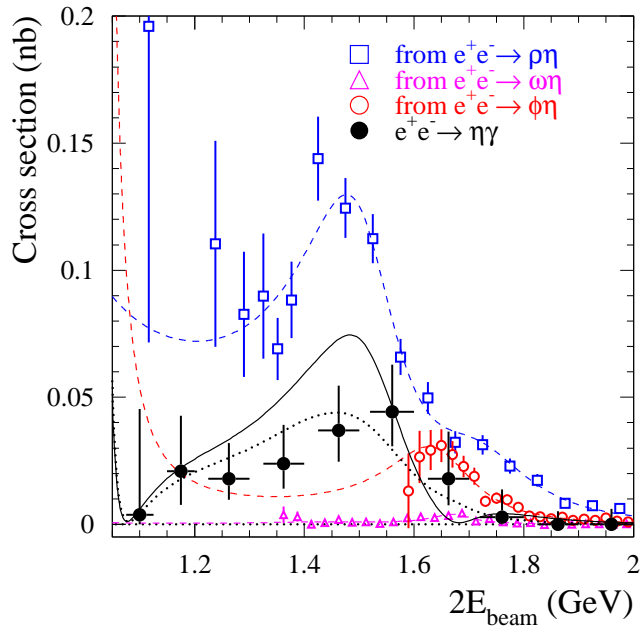}
\caption{ \label{crs2}
The $e^+e^-\to\eta\gamma$ cross section measured in this work 
({\LARGE $\bullet$}) in comparison with the cross 
sections calculated using VMD from data on the processes 
$e^+e^-\to\rho\eta$ ($\square$), $e^+e^-\to\omega\eta$ ($\bigtriangleup$) and
$e^+e^-\to\phi\eta$ ({\LARGE $\circ$}). The dashed curves are results of
the approximation of these cross sections by Eq.~(\ref{parcrs}). The 
solid curve is the total cross section taking into account the interference 
of the isovector and isoscalar amplitudes. The dotted curve is the total cross
section calculated using modified excited-state amplitudes (see the text).
}
\end{figure}

The calculated contributions to the $e^+e^-\to \eta\gamma$ cross section 
are shown in Fig.~\ref{crs2} in comparison with the cross section measured 
in this work. For the $e^+e^-\to\rho\eta$ cross section,  experimental data 
from \cite{rhoeta,BABARomegapipi} were used, while for the $e^+e^-\to\omega
\eta$ and $e^+e^-\to\phi\eta$ cross sections data from  
Refs.~\cite{omegaeta,phieta}, respectively. It is seen that the 
contribution from the $\omega$-like excited states is small. The VMD 
calculation confirms the predictions of the potential quark 
model~\cite{thpred} that the dominant contribution comes from the isovector 
and  $s\bar{s}$ excited resonances. The calculated isovector cross section, 
even considering its 50\% uncertainty, is significantly higher than the 
measured $e^+e^-\to \eta\gamma$ cross section. The difference can be 
reduced by a destructive interference of the isovector and $s\bar{s}$
amplitudes.

In order to calculate the interference effects, the cross sections derived
by using Eq.~(\ref{VMD}) are fit with Eq.~(\ref{parcrs}). For the isovector
cross section, contributions  from the
$\rho(770)$, $\rho(1450)$ and $\rho(1700)$ only are considered.
The isoscalar $s\bar{s}$ ($n\bar{n}$) cross section is described by 
a sum of the contributions of the $\phi$($\omega$)-like states. 

Parameters (the mass, width and the peak cross section) for
the $\rho(770)$, $\omega(782)$ and $\phi(1020)$ states are fixed at their 
PDG values~\cite{pdg}. Parameters of the excited resonances $\rho(1450)$, 
$\omega(1650)$ and $\phi(1680)$, which give the dominant contribution to the 
corresponding cross sections, are free fit parameters. The phase 
difference between the ground state and the first excitation is chosen to 
be $180^\circ$. Parameters of the $\rho(1700)$ and $\omega(1420)$ resonances
are fixed at their PDG values~\cite{pdg}. For the phases of the $\omega(1650)$ 
and $\rho(1700) $ two  options $0^\circ$ and $180^\circ$ were checked. In the 
case of the $\rho(1700)$, the best fit is obtained when the phase 
$\varphi_{\rho(1700)}=\varphi_{\rho}+180^\circ$, while for the $\omega(1650)$ 
at the phase $\varphi_{\omega(1650)}=\varphi_{\omega}$. Note that because
of the relative smallness of the $\rho(1700)$ and $\omega(1650)$ 
contributions, choice of these phases has little influence on the size and 
shape of the total cross section. The masses and widths of the $\rho(1450)$, 
$\omega(1650)$ and $\phi(1680)$ found in the fit  are consistent
with the PDG values~\cite{pdg}. The fit results 
are shown in Fig.~\ref{crs2} by the dashed curves.

The isovector and isoscalar $n\bar{n}$ and $s\bar{s}$ amplitudes obtained 
from the fit are combined with phases 
$\varphi_{\omega}=\varphi_{\rho}$ and $\varphi_{\phi}=\varphi_{\rho}+
180^\circ$. The resulting total cross section shown in Fig.~\ref{crs2} by 
the solid curve is in rather good agreement (considering the 50\% 
uncertainty of the VMD calculation) with the measured cross section. 
A relatively small modification of the isovector and $s\bar{s}$ amplitudes 
(the $\rho(1450)$ and $\rho(1700)$ amplitudes are reduced by  15\%, and 
the $\phi(1680)$ amplitude is increased by 25\%) significantly improves this 
agreement. The cross section obtained after this modification
is shown in Fig.~\ref{crs2} by the dotted curve.

Thus, our analysis in this section confirms that the dominant contribution 
to the cross section of $e^+e^- \to \eta\gamma$ in the energy range 
1.1--2.0~GeV comes from the radiative decays of the two excited vector
resonances $\rho(1450)$ and $\phi(1680)$.
 
\section{Conclusion}
The cross section for the process  $e^+e^- \to \eta\gamma$ has been measured
in the center-of-mass energy range from 1.07 to 2.00~GeV 
with the SND detector at the VEPP-2000 $e^+e^-$ collider. 
Above 1.4 GeV, the cross section for this process has been measured for 
the first time. About 30 $e^+e^- \to \eta\gamma$ events
detected at c.m. energy above 1.15 GeV cannot be 
explained within the VMD model with the $\rho(770)$, $\omega(782)$ 
and $\phi(1020)$ mesons only. We interpret these events as an observation of 
radiative decays of excited vector states into $\eta\gamma$.

From the combined analysis of the $e^+e^- \to \eta\gamma$ data
obtained in this work and the data on the cross sections for 
$e^+e^- \to \rho\eta$, $e^+e^- \to \omega\eta$ and 
$e^+e^- \to \phi\eta$ we make the conclusion that the main contribution 
to the $e^+e^- \to \eta\gamma$ cross section above 1.1~GeV comes from the 
decays of the two excited meson states $\rho(1450)$ and $\phi(1680)$. For the 
processes $e^+e^- \to \rho(1450)\to \eta\gamma$ and $e^+e^- \to \phi(1680)\to 
\eta\gamma$, the following cross section values at the resonance peaks have
been obtained:
\begin{eqnarray}
\sigma_{\rho(1450)\to\eta\gamma} = 57 \pm 10 \pm 7\mbox{ pb},\nonumber \\
\sigma_{\phi(1680)\to\eta\gamma} = 52 \pm 17 \pm 15\mbox{ pb}.
\end{eqnarray}
These cross sections can be compared with the predictions of the quark model.
In Ref.~\cite{thpred} the following values for the partial
decay widths were obtained: $\Gamma_{\rho(1450)\to\eta\gamma}\approx
\Gamma_{\phi(1680)\to\eta\gamma}\approx 100$ keV.
Using PDG values of the widths of the resonances and rough estimates of the 
total production cross sections of the $\rho(1450)$ (60~nb from the sum of 
the cross sections of $e^+e^-\to \pi^+\pi^-\pi^0\pi^0$ and
$e^+e^-\to \pi^+\pi^-\pi^+\pi^-$~\cite{druzhinin}), and $\phi(1680)$
(13~nb from the sum of the cross sections $e^+e^-\to K\bar{K}^\ast$ 
and $e^+e^-\to \phi\eta$~\cite{phieta}) resonances in $e^+e^-$ annihilation, 
we can estimate the cross sections 
$\sigma_{\rho(1450)\to\eta\gamma} \approx 15$~pb and 
$\sigma_{\phi(1680)\to\eta\gamma} \approx 10$~pb. It is seen that the 
decay widths of 
$\rho(1450)\to\eta\gamma$ and $\phi(1680)\to\eta\gamma$ obtained in 
Ref.~\cite{thpred} are too 
small to explain the observed $e^+e^- \to \eta\gamma$ cross section.

We thank S. I. Eidelman for useful discussions.
This work is supported by the Ministry of Education and Science of 
the Russian Federation, the Russian Federation 
Presidential Grant for Scientific Schools NSh-5320.2012.2, 
RFBR (grants 12-02-01250, 12-02-00065, 13-02-00375, 13-02-00418), 
Grant 14.740.11.1167 from the Federal Program ``Scientific and Pedagogical
Personnel of Innovational Russia''.

\end{document}